\documentclass[prl,amsmath,amsfonts,amssymb,letterpaper,superscriptaddress,twocolumn,showpacs]{revtex4}
\pdfoutput=1
\usepackage{bm}
\usepackage{graphicx}
\usepackage{hyperref}
\usepackage{subfigure}

\hypersetup{
colorlinks=true,
    citecolor=blue,
    pdfborder={0 0 0},
}

\newcommand\arrow\vec

\def\vec#1{\bm{#1}}
\DeclareMathOperator{\sech}{sech}

\def\negspace{\!}
\def\lrsub#1#2#3{{\vphantom{#1}}_{#2} \negspace {#1} \negspace {\vphantom{#1}}_{#3}}

\def\bra#1{\left\langle {#1} \right\rvert}
\def\ket#1{\left\lvert {#1} \right\rangle}

\def\inprod#1#2{\left\langle {#1} | {#2} \right\rangle}

\def\inprodsubsub#1#2#3#4{\lrsub {\inprod{#1}{#2}} {#3} {#4}}

\def\pqinprod#1#2{\inprodsubsub{#1}{#2} p q}
\def\qpinprod#1#2{\inprodsubsub{#1}{#2} q p}
\def\pinprod#1#2{\inprodsubsub{#1}{#2} p p}
\def\qinprod#1#2{\inprodsubsub{#1}{#2} q q}
\def\pqbraket\pqinprod
\def\qpbraket\qpinprod
\def\pbraket\pinprod
\def\qbraket\qinprod

\def\outprod#1#2{\ket {#1}\!\bra {#2}}

\def\1{I}

\def\v0{{\bvec 0}}

\begin{document}

\title{Beating the One-half Limit of Ancilla-free Linear Optics Bell Measurements}

\author{Hussain A. Zaidi}
\email{haz4z@virginia.edu}
\affiliation{Optical Quantum Information Theory Group, Max Planck Institute for the Science of Light, G\"unther-Scharowsky-Str. 1/Bau 24, 91058 Erlangen, Germany}

\author{Peter van Loock}
\affiliation{Optical Quantum Information Theory Group, Max Planck Institute for the Science of Light, G\"unther-Scharowsky-Str. 1/Bau 24, 91058 Erlangen, Germany}
\affiliation{Institute of Physics, Johannes-Gutenberg Universit\"at Mainz, Staudingerweg 7, 55128 Mainz, Germany}


\date\today

\begin{abstract}
We show that optically encoded two-qubit Bell states can be unambiguously discriminated with a success probability of more than $50\%$ in both single-rail and dual-rail encodings by using active linear-optical resources that include Gaussian squeezing operations.
%single mode squeezing (an operation that is effectively linear), beam splitters and photon number resolving detectors %(PNRDs).
These results are in contrast to the well-known upper bound of $50\%$ for unambiguous discrimination of dual-rail Bell states using passive, static linear optics and arbitrarily many vacuum modes.
We present experimentally feasible schemes that improve the success probability to 64.3\% in dual-rail and to 62.5\% in single-rail for a uniform random distribution of Bell states. 
%, and PNRDs.
Conceptually, this demonstrates that neither interactions that induce nonlinear mode transformations (such as Kerr interactions) nor auxiliary entangled photons are required to go beyond the one-half limit. We discuss the optimality of our single-rail scheme, and talk about an application of our dual-rail scheme in quantum communication.
\end{abstract}

\maketitle

%%%%%%%%%%%%%%%%%%%%%%%%%%%%%%%%%%
% Introduction
%%%%%%%%%%%%%%%%%%%%%%%%%%%%%%%%%%
%\section{I. Introduction}
\emph{Introduction---} Bell measurements, jointly projecting two qubits onto the so-called Bell basis, constitute a crucial step in many quantum computation and communication protocols, including dense coding \cite{mattle1996}, quantum repeaters \cite{duan2001}, and teleportation-based quantum computation \cite{gottesman1999, knill2001}. The two most common encodings for optical Bell states are the single-rail (SR) and the dual-rail (DR) encodings \cite{furusawa2011}. An important result with regards to Bell measurements (BMs) is the impossibility of deterministic unambiguous DR BMs using passive linear optics \cite{kok2007}, 
even when arbitrarily many
auxiliary photons, photon-number-resolving detectors (PNRDs), and dynamical (conditionally changing) networks are available \cite{luetkenhaus1999}. Quantitatively, the success probability for unambiguous DR BMs using a static linear network aided by vacuum modes and PNRDs was shown to be tightly bounded from above by $50\%$ \cite{calsamiglia2001}. In the supplementary section of this work, we give a simple proof showing that the upper bound of $50\%$ also holds for SR BMs, even with the inclusion of dynamical networks \footnote{Converting between SR and DR encodings requires nonlinear optics, which means that there is no intuitive way of inferring the success probability in SR given that in DR, and vice versa. Hence, throughout this paper, we treat DR and SR encodings separately.}. A fully deterministic BM requires at least a cubic Hamiltonian \footnote{With regards to the possibility of a deterministic BM with the help
of arbitrary quadratic Hamiltonians together with PNRDs, auxiliary photons, and conditional dynamics, no-go theorems
can be derived for both SR and DR encodings by extending the formalism presented in \cite{vanloock2004} to the case of arbitrary linear transformations including squeezing. The relevant calculations have been omitted in the interest of space, but are a straightforward extension of the formalism of Ref. \cite{vanloock2004}.} (which is highly inefficient in practice \cite{kim2001}), unless one relies on embedded Bell-state analysis \cite{barreiro2008}. Current proposals for going past the $50\%$ upper bound without using experimentally challenging nonlinearities rely on using entangled photon ancilla states (which are generally expensive and probabilistic to create) and a sufficiently large interferometer to combine the signal and ancilla modes \cite{grice2011}, \footnote{Technically, the scheme given in Ref. \cite{grice2011} is for DR, but an extension of this scheme also works in SR, and will be given elsewhere.}. Similar to Ref. \cite{knill2001}, BMs in these proposals are made near-deterministic in the limit of asymptotically large ancilla states, but without the need for conditional dynamics.

In light of the above facts, the motivation for this work is threefold. First, from a theoretical point of view, while an upper bound on the success probability of unambiguous Bell discrimination using {\it passive} linear optics and auxiliary vacuum modes has been established and well celebrated, no such upper bound has been shown to exist for
{\it active} linear-optical circuits that include additional single-mode squeezers and correspond to arbitrary quadratic Hamiltonians \cite{braunstein2005_77}
(and hence arbitrary linear transformations of the mode operators \footnote{As mentioned earlier, we only know that a fully deterministic BM requires at least a cubic Hamiltonian.}).
Even though there is a renewed interest in the area of BMs \cite{grice2011,pavicic2011}, there exists a gap in our understanding of what is possible between passive linear optics and cubic nonlinearities \footnote{Ref. \cite{pavicic2011}, which claimed to show near-deterministic BMs (with an arbitrarily small ambiguity) in DR by using passive linear optics and vacuum modes, was later retracted because of erroneous results.}. Second, historically, squeezing has typically been used in {\it continuous-variable} quantum computing and information processing \cite{braunstein2005_77, weedbrook2012}. We, however, want to explore the use of squeezing in \emph{discrete-variable} BMs, opening up the possibility of combining continuous-variable and discrete-variable toolkits for enhanced quantum information processing \cite{furusawa2011}. Finally, from an experimental point of view, squeezing low-photon-number states has become an achievable feat in the last few years
%\cite{yoshikawa2007}.
\cite{miwa2012}, which warrants an exploration into its potential for quantum information processing.
Note that in Ref. \cite{miwa2012}, optical squeezing has been promoted from an offline experimental resource to a controllable online operation, just as is needed for our purposes.

We show in this paper that beating the $50\%$ bound on the success probability of unambiguous BM is possible by using single-mode squeezers and beam splitters. 
%The setups presented in this paper yield $62.5\%$ and $64.3\%$ success probability in SR and DR, respectively. 
Our schemes are free of ancilla photons and vacuum modes. Hence, the static interferometric networks needed are no larger than the signal-mode space (i.e., four modes in DR and two modes in SR).
We first present a scheme with a success probability of $64.3\%$ for DR unambiguous BM (while we do not know if this is optimal). We then consider SR BMs using an arbitrary two-mode network of beam splitters and squeezers characterized by real parameters. We show that for equal squeezing in the two modes, there exists an entire class of interferometers that yield $62.5\%$ success probability. We then look at experimental considerations and possible applications of our schemes. In the supplementary section, we discuss numerical results that show that it is not possible to beat $62.5\%$ with a real two-mode active linear network (i.e., without phase shifts) for a squeezing of up to $8.686$ dB. 

\begin{figure}[htb]
\begin{center}
\begin{tabular}{c}
\includegraphics[width= .7 \columnwidth]{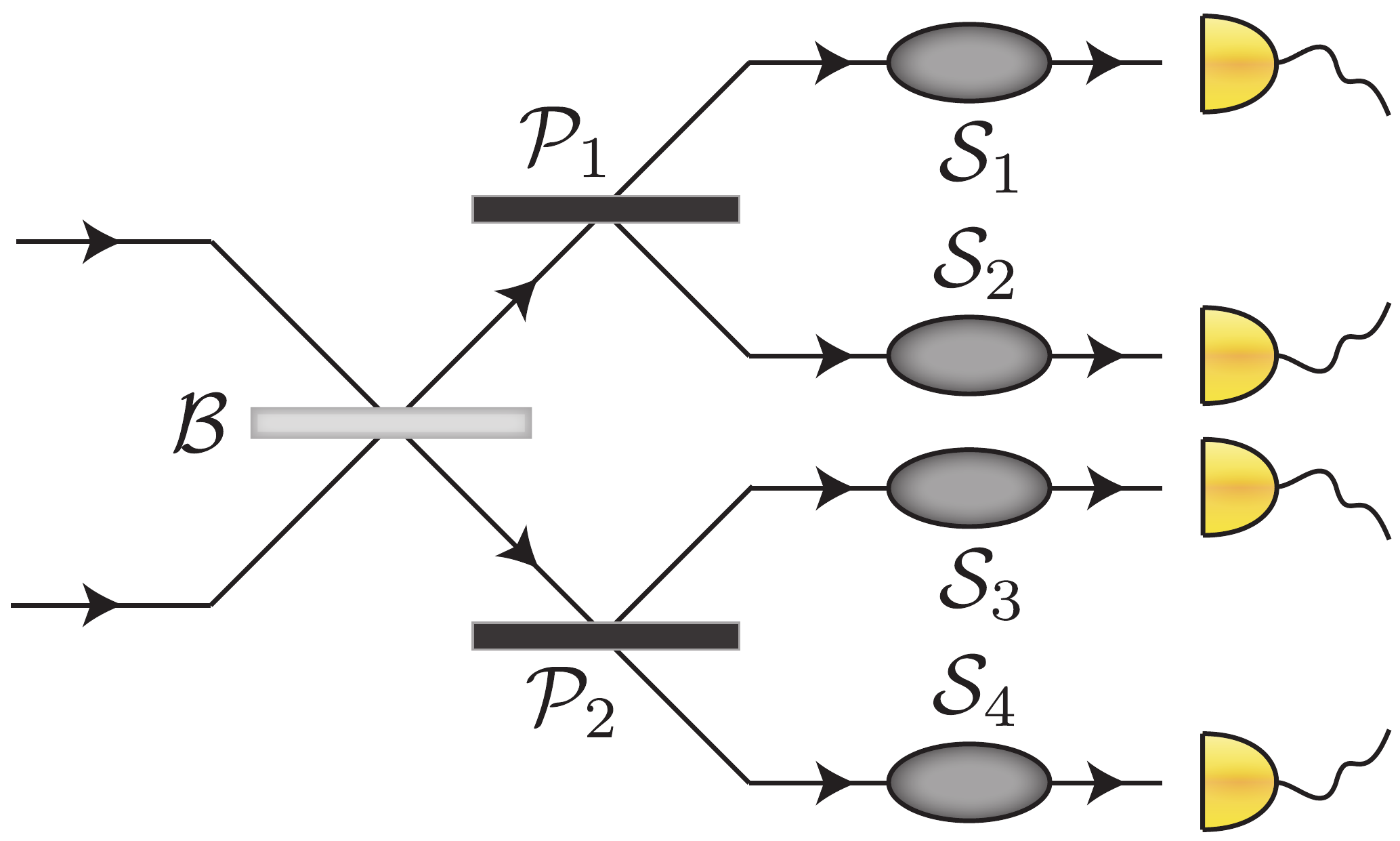}
\end{tabular}
\end{center}
\caption{Scheme for DR Bell discrimination that yields a success probability of $64.2\%$. The Bell states pass through a balanced beam splitter ($\mathcal{B}$), two polarizing beam splitters ($\mathcal{P}_1$ and $\mathcal{P}_2$), and four squeezers ($\mathcal{S}_1$ to $\mathcal{S}_4$) with a squeezing of $5.7195$ dB each. The output is sent to PNRDs.}
\label{fig:DRScheme}
\end{figure}

\emph{Dual-Rail Bell Discrimination---} Let us start with DR Bell states in polarization basis, given by
\begin{subequations}
\begin{align}
\ket{\psi^\pm}=\frac{1}{\sqrt{2}}\left(\ket{HV}\pm\ket{VH}\right),\\
\ket{\phi^\pm}=\frac{1}{\sqrt{2}}\left(\ket{HH}\pm\ket{VV}\right),
\end{align}
\end{subequations}
where $H$ and $V$ stand for horizontal and vertical polarizations, respectively. Consider the setup shown in Fig. [\ref{fig:DRScheme}] with Bell states as the input.
%, in which the Bell states pass through a beam splitter, a pair of polarization beam splitters, and finally through four equal-strength squeezers before being detected by PNRDs.
%------------
%------------
We choose a balanced beam splitter whose action on the mode operator vector $(a_1^\dagger\mbox{ }a_2^\dagger)^T$ is represented by the matrix
$\frac{1}{\sqrt{2}} \begin{pmatrix}
		1		& i\\
		i		&1
	\end{pmatrix}$.
%\begin{align}
%\frac{1}{\sqrt{2}} \begin{pmatrix}
%		1		& i\\
%		i		&1
%	\end{pmatrix}.
%\end{align}
After passing through the beam splitter and the polarizing beam splitters (where vertically polarized photons are reflected), the Bell states transform into
\begin{subequations}
\label{eq:DRPhiAfterPBS}
\begin{align}
&\ket{\psi^+}\rightarrow\frac{i}{\sqrt{2}}\left(\ket{1100}+\ket{0011}\right),\\
&\ket{\psi^-}\rightarrow\frac{1}{\sqrt{2}}\left(\ket{1010}-\ket{0101}\right),\\
&\ket{\phi^\pm}\rightarrow\frac{i}{2}\left(\ket{2000}+\ket{0002}\pm\ket{0200}\pm\ket{0020}\right).
\end{align}
\end{subequations}
We will disregard global phases from now on. Notice that at this point the states $\ket{\psi^\pm}$ would be perfectly distinguishable if we were to detect four-mode photon-number patterns, while the states $\ket{\phi^\pm}$ would be indistinguishable. Single-mode squeezers add photons in pairs, thereby preserving the even and odd parity of the number of photons in each output mode. Hence, the states $\ket{\psi^\pm}$ remain perfectly distinguishable after applying squeezing. For this reason, we only concern ourselves with the states $\ket{\phi^\pm}$ in the following analysis.

Now consider the squeezing operator $\mathcal{S}$ given by $\exp{[r (a^2-a^{\dagger^{2}})/2]}$. After normal ordering, we can write its effect on the relevant Fock states $\ket{0}$, $\ket{1}$, and $\ket{2}$ as
\begin{subequations}
\label{eq:Squeezing}
\begin{align}
\mathcal{S}\ket{0}=&\sqrt{\sech r}\exp{(-\tanh r a^{\dagger^2}/2)}\ket{0},\\
\mathcal{S}\ket{1}=&(\sech r)^{3/2}\exp{(-\tanh r a^{\dagger^2}/2)}\ket{1},\\
\mathcal{S}\ket{2}=&\sqrt{\sech r/2}\tanh r\exp{(-\tanh r a^{\dagger^2}/2)}\ket{0}\nonumber\\
 &\mbox{} +(\sech r)^{5/2}\exp{(-\tanh r a^{\dagger^2}/2)}\ket{2}.
\end{align}
\end{subequations}
Each Bell state is transformed into an infinite sum of photonic four-mode states when it is passed through the squeezers. After some algebra, we can identify the coefficients for the different photonic states (the coefficients for the states $\ket{2000}$, $\ket{4246}$, etc.). One possible strategy for unambiguous discrimination, then, relies on choosing a squeezing value where some of these coefficients are zero for $\ket{\phi^+}$, but non-zero for $\ket{\phi^-}$, so that some outputs occur unambiguously for $\ket{\phi^-}$. Eq. (\ref{DRPhiFirstOrder}) below shows the output from the squeezers up to two-photon terms:  
\begin{align}
\label{DRPhiFirstOrder}
\ket{\phi^\pm}\rightarrow&\frac{\alpha^\pm}{\sqrt{2}}\ket{0000}-\frac{1}{2}(\tanh r\alpha^\pm-\sech^4 r)(\ket{2000}+\ket{0002})\nonumber\\
&-\frac{1}{2}(\tanh r\alpha^\pm\mp\sech^4 r)(\ket{0200}+\ket{0020}),
\end{align}
where $\alpha^\pm=\tanh r\sech^2 r(1\pm 1)$. 
%In general, lower photon number outputs have larger coefficients, which makes it a good strategy to make lower photon number terms distinguishable. For example, it is better to distinguish the $\ket{2000}$ term than the $\ket{4000}$ term.
The complete output is presented in the supplementary section, but we can already see that the vacuum output is unique to the input $\ket{\phi^+}$, leading to an above-50\% success probability of BM for any non-zero value of squeezing. By picking $r=0.6585$ (a squeezing of $5.7195$ dB \footnote{The conversion from the unitless parameter $r$ to squeezing in dB is $-10\log_{10} (\exp{(-2r)})$.}), so that $\tanh r\alpha^+-\sech^4 r=0$, we can make all the two-photon output terms unique to $\ket{\phi^-}$. At this value, the vacuum coefficient in $\ket{\phi^+}$ is also at its maximum. Hence, the most significant unambiguous output terms for our input Bell states $\ket{\phi^\pm}$ (denoted by $\ket{\phi^\pm}_{unique}$) are given by:
\begin{align}
\ket{\phi^+}_{unique}\sim & +0.5443\ket{0000},\nonumber\\
\ket{\phi^-}_{unique}\sim & +0.2222\ket{2000}-0.2222\ket{0200}\nonumber\\
& -0.2222\ket{0020}+0.2222\ket{0002}.
\end{align}
%From the above expression we can see that the success probability of unambigious discrimination is $29.63\%$ for $\ket{\phi^+}$ and $19.75\%$ for $\ket{\phi^-}$.
As a by-product of the above condition, a number of higher-photon-number terms also become unambiguous (i.e., only show up for $\ket{\phi^+}$). Adding up the success probabilities from these higher-order terms, we obtain a total probability of $37.48\%$ and $19.75\%$ for the inputs $\ket{\phi^+}$ and $\ket{\phi^-}$, respectively.
%The fact that the success probabilities for $\ket{\phi^+}$ and $\ket{\phi^-}$ in dual-rail are unequal is surprising and in contrast to the single-rail case for which the success probabilities are symmetric.
Combining this with deterministic discrimination of $\ket{\psi^\pm}$, and assuming equal input probability for each of the Bell states, the network shown in Fig. [\ref{fig:DRScheme}] yields an overall success probability of $64.3\%$.

%\section{III. Success Probability Using a Real Two Mode Network with Squeezing}
\emph{Single-Rail Bell Discrimination---} In SR, similar to the DR case, a scheme relying on a balanced beam splitter followed by two single-mode squeezers, each with a squeezing of $6.2696$ dB ($r=0.7218$) produces a success probability of $62.5\%$. However, here we shall consider a more general setup: a two-mode linear circuit parametrized by real reflectivity and squeezing parameters. The equal-squeezing scheme is later discussed as a special case of this general setup.
 %We will later discuss our numerical results for such general networks to shed some light on the optimality of a two-mode active linear network.

\begin{figure}[htb]
\begin{center}
\begin{tabular}{c}
\includegraphics[width= .8 \columnwidth]{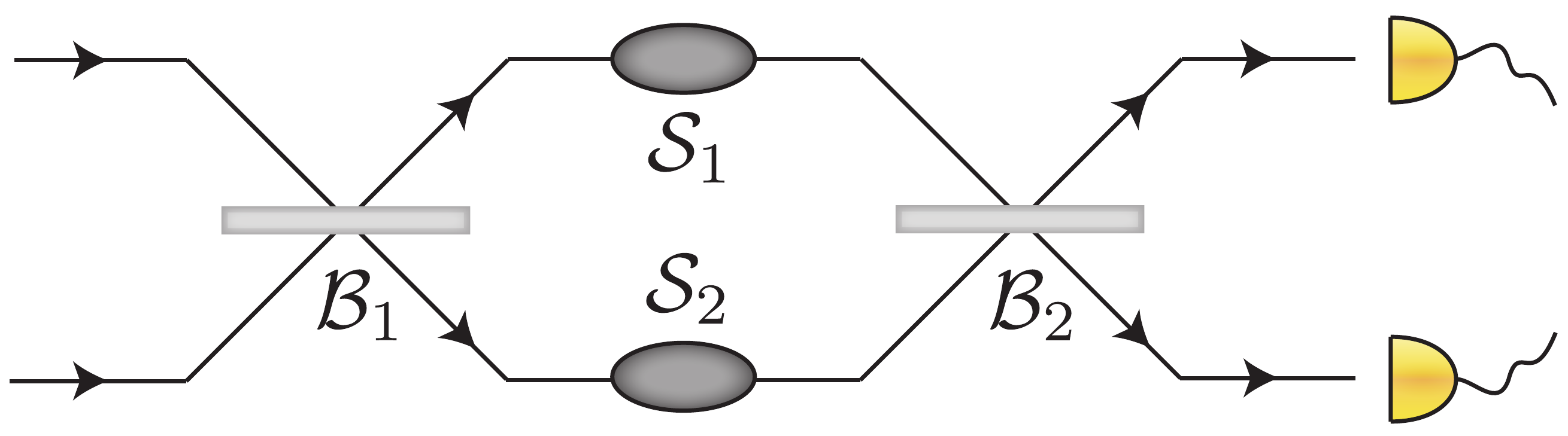}
\end{tabular}
\end{center}
\caption{In SR, we consider the Bloch-Messiah reduction of a general two-mode squeezing network. The Bell states are passed through a beam splitter $\mathcal{B}_1(\theta_1)$, followed by squeezers $\mathcal{S}_1(r_1)$ and $\mathcal{S}_2(r_2)$, and finally through a second beam splitter $\mathcal{B}_2(\theta_2)$; $\theta_1$ and $\theta_2$ are the reflectivity parameters, and $r_1$ and $r_2$ are the squeezing parameters. The output is sent to PNRDs. We only discuss networks with real parameters.}
\label{fig:SRScheme}
\end{figure}

Bell states in SR are given by:
\begin{subequations}
\begin{align}
\ket{\psi^\pm}=\frac{1}{\sqrt{2}}(\ket{10}\pm\ket{01}),\\
\ket{\phi^\pm}=\frac{1}{\sqrt{2}}(\ket{00}\pm\ket{11}).
\end{align}
\end{subequations}
Using the Bloch-Messiah reduction \cite{braunstein2005_71}, an arbitrary two-mode squeezing network can be decomposed into a combination of beam splitters and single-mode squeezers, as shown in Fig. [\ref{fig:SRScheme}].
%------------
%------------
%Subscripts 1 and 2 signify the mode number, except for when they are used with beam splitters in which case they signify the first or the second beam splitter.
The effect of the first beam splitter $\mathcal{B}_1$ on the operator vector $(a_1^\dagger \mbox{ } a_2^\dagger)^T$ is given by the matrix $\begin{pmatrix}
		\cos \theta_1		& \sin \theta_1\\
		-\sin \theta_1		&\cos \theta_1
	\end{pmatrix}$,
%\begin{align}
%B_1= \begin{pmatrix}
%		\cos \theta_1		& \sin \theta_1\\
%		-\sin \theta_1		&\cos \theta_1
%	\end{pmatrix},
%\end{align}
and that of the second beam splitter is given by the substitution $\theta_1\rightarrow\theta_2$. Applying the beam splitters and squeezers sequentially on the input Bell states, the states $\ket{\psi^\pm}$ are transformed into:
%\mathcal{B}_2&\mathcal{S}_2\mathcal{S}_1\mathcal{B}_1
\begin{align}
\ket{\psi^\pm}\rightarrow &\sqrt{\frac{\sech r_1 \sech r_2}{2}}\exp{(xa_1^{\dagger^2}+ya_2^{\dagger^2}+za_1^\dagger a_2^\dagger)}\nonumber\\
&\times\left(\Omega^\pm\ket{10}+\omega^\pm\ket{01}\right),\label{eq:PsiPMOutputSR}
\end{align}
where $x=-\tanh r_1\cos^2\theta_2/2 -\tanh r_2\sin^2\theta_2/2$, $y=-\tanh r_1\sin^2\theta_2/2 -\tanh r_2\cos^2\theta_2/2$, $z=\cos\theta_2 \sin\theta_2 (-\tanh r_1 +\tanh r_2)$, and
\begin{subequations}
\label{eq:PsiConstraints}
\begin{align}
%x\hspace{9pt}=&-\tanh r_1\cos^2\theta_2/2 -\tanh r_2\sin^2\theta_2/2\\
%y\hspace{9pt}=&-\tanh r_1\sin^2\theta_2/2 -\tanh r_2\cos^2\theta_2/2\\
%z\hspace{9pt}=&\cos\theta_2 \sin\theta_2 (-\tanh r_1 +\tanh r_2)\\
\Omega^\pm=&\sech r_1\cos\theta_2(\cos\theta_1\mp\sin\theta_1)\nonumber\\
 &-\sech r_2\sin\theta_2(\sin\theta_1\pm \cos\theta_1),\\
\omega^\pm=&\sech r_1\sin\theta_2(\cos\theta_1\mp\sin\theta_1)\nonumber\\
 &+\sech r_2\cos\theta_2(\sin\theta_1\pm \cos\theta_1).
\end{align}
\end{subequations}
The case of passive linear optics is obtained by setting $r=0$. Specifically, $\ket{\psi^+}\rightarrow\ket{01}$ and $\ket{\psi^-}\rightarrow\ket{10}$ can be recovered by suitably choosing $\theta_1$ and $\theta_2$, e.g., $\theta_1=0$ and $\theta_2=\pi/4$. By writing out the first few terms in Eq. (\ref{eq:PsiPMOutputSR}), it can be seen that for an apparatus that measures photon numbers in each mode, at least one of the four conditions ($\Omega^+=0$, $\Omega^-=0$, $\omega^+=0$ or $\omega^-=0$) must be satisfied to unambiguously discriminate between $\ket{\psi^\pm}$ with a non-zero probability. If we choose to ignore the above constraints, we may distinguish the states $\ket{\phi^\pm}$ at most deterministically, but that will yield an overall success probability of $50\%$, which is no better than that achieved with a passive linear network. Hence, satisfying one of these conditions is necessary to go beyond the passive linear upper bound of $50\%$.  
%Each condition (or a combination of these conditions) specifies some or all of the parameters of the optical network.

%Having identified the necessary constraints for going beyond the passive linear upper bound, we look at the output for the input $\ket{\phi^\pm}$:
We now look at the output for the states $\ket{\phi^\pm}$:
%\mathcal{B}_2&\mathcal{S}_2\mathcal{S}_1\mathcal{B}_1
\begin{align}
\ket{\phi^\pm}\rightarrow&\sqrt{\frac{\sech r_1 \sech r_2}{2}}\exp{(xa_1^{\dagger^2}+ya_2^{\dagger^2}+za_1^\dagger a_2^\dagger)}\nonumber\\
&\times\left(\gamma^\pm\ket{00}+\rho^\pm\ket{02}-\rho^\pm\ket{20}+\zeta^\pm\ket{11}\right),\label{eq:PhiPMOutputSR}
\end{align}
where
\begin{subequations}
\label{eq:PhiCoefficients}
\begin{align}
\gamma^\pm=&1\mp\sin 2\theta_1(\tanh r_1 - \tanh r_2)/2,\\
\rho^\pm=&\pm 1/\sqrt{2}\left(\cos 2\theta_1\sin 2\theta_2\sech r_1\sech r_2\right.\nonumber\\
&-\left. \sin 2\theta_1(\sin^2\theta_2 \sech^2 r_1 - \cos^2\theta_2\sech^2 r_2)\right),\\
\zeta^\pm=&\pm \cos 2\theta_1\cos 2\theta_2 \sech r_1\sech r_2\nonumber\\
&\mp \sin 2\theta_1\sin 2\theta_2 (\sech^2 r_1+\sech^2 r_2)/2.
\end{align}
\end{subequations}
Eqs. (\ref{eq:PsiPMOutputSR}) and (\ref{eq:PhiPMOutputSR}) show that $\ket{\psi^\pm}$ always result in an odd number of photons in the output, while $\ket{\phi^\pm}$ result in an even number. Hence, we only have to distinguish $\ket{\psi^+}$ from $\ket{\psi^-}$, and $\ket{\phi^+}$ from $\ket{\phi^-}$.
Eqs. (\ref{eq:PsiPMOutputSR}) and (\ref{eq:PhiPMOutputSR}) allow us to numerically analyze the case of a general two-mode network in the supplementary section. In the following, we analytically consider the case of equal squeezing ($r_1=r_2$), for which we achieve a success probability of 62.5\%.

%\subsection{Network with Equal Squeezing}
\emph{Single-Rail, Equal Squeezing---} For $r_1=r_2=r$, $\Omega^+=0\Rightarrow \omega^-=0$, and $\Omega^-=0\Rightarrow \omega^+=0$, as can be seen from Eq. (\ref{eq:PsiConstraints}). Hence, imposing any of the four conditions renders the states $\ket{\psi^\pm}$ completely distinguishable, as evident from Eq. (\ref{eq:PsiPMOutputSR}). Further, we only need to look at the constraints $\Omega^\pm=0$ to infer the success probabilities for $\omega^\pm=0$. From Eqs. (\ref{eq:PsiConstraints}) and (\ref{eq:PhiCoefficients}), we have
\begin{align}
%&\Omega^+=0\Rightarrow \begin{cases} \rho^\pm=\pm \sech^2 r /\sqrt{2} %\mbox{ , }\; \zeta^\pm=0\smallskip\\ \cos^2\theta_2=(1+\sin 2\theta_1)/2 \end{cases}\label{eq:OmegaPConstraint}
&\Omega^\pm=0\Rightarrow \begin{cases}  \cos^2\theta_2=(1\pm\sin 2\theta_1)/2\mbox{ , }\smallskip\\ \rho^+=\pm \sech^2 r /\sqrt{2} \mbox{ , }\; \zeta^+=0\smallskip\mbox{ , }\\ \rho^-=\mp \sech^2 r /\sqrt{2} \mbox{ , }\; \zeta^-=0\mbox{ . }
 \end{cases}\label{eq:OmegaPConstraint}
\end{align}
%\begin{align}
%&\Omega^-=0\Rightarrow \begin{cases} \rho^\pm=\mp \sech^2 r /\sqrt{2} %\mbox{ , } \; \zeta^\pm=0 \smallskip\\ \cos^2\theta_2=(1-\sin 2\theta_ 1)/2 \end{cases}\label{eq:OmegaMConstraint}
%\end{align}
Combined with $x=y=-\tanh r/2$, $z=0$, and $\gamma^\pm=1$, we see that after imposing $\Omega^\pm=0$ the coefficients for $\ket{\phi^\pm}$ are independent of $\theta_1$ and $\theta_2$. This implies that each value of $r$ gives an entire class of networks with the same unambiguous-discrimination success probability. For example, a network with a balanced beam splitter after equal-strength squeezers gives the same success probability as the one with the order reversed. Further, setting $r=0$ gives us an entire class of passive networks for which $\ket{\psi^\pm}$ are distinguishable and $\ket{\phi^\pm}$ are indistinguishable. As a final observation, since $\Omega^+=0$ and $\Omega^-=0$ give only an immaterial sign difference in $\rho^\pm$, we are free to choose either condition for determining the success probability. In the following, we have chosen $\Omega^+=0$.

We can now evaluate Eq. (\ref{eq:PhiPMOutputSR}) with the above simplifications to write the output $\ket{\phi^\pm}$ analytically in Fock space. The complete output is presented in the supplementary section, but the following truncated form highlights the main aspects of the calculation (similar to the DR case):
%\mathcal{B}_2\mathcal{S}_2\mathcal{S}_1\mathcal{B}_1
\begin{align}
\ket{\phi^\pm}\rightarrow &\frac{\sech r}{\sqrt{2}}\ket{00}+\frac{\sech r}{2}(-\tanh r\mp\sech^2 r)\ket{20}\nonumber\\
&+\frac{\sech r}{2}(-\tanh r \pm\sech^2 r)\ket{02}+\mbox{...},
\end{align}
where we have only kept terms up to two photons.
While the vacuum output is always ambiguous, the states $\ket{20}$ and $\ket{02}$ can be made unambiguous at $r=0.7218$ (a squeezing of $6.2696$ dB). Adding up the success probabilities from higher-order unambiguous terms yields a success probability of $25.03\%$ for each of the states $\ket{\phi^\pm}$, resulting in an overall probability of $62.5\%$. We find numerically that for a phase-free and at most $8.686$ dB squeezed, but otherwise arbitrary, two-mode squeezing network, $62.5\%$ is indeed the best efficiency (discussion given in the supplementary part).

\emph{Experimental Success Probabilities---} To take into account the experimental upper limit on the photon-number resolution of the PNRDs, we calculated {\it effective} success probabilities for equal-squeezing schemes considering PNRD resolutions of $2$, $5$, and $10$ photons (while still assuming unit detection efficiency). Table [\ref{tab:ExpSuccessProb}] shows that almost all the benefits of the presented schemes can be harnessed with PNRDs that can resolve up to $10$ photons \footnote{Notice that for a resolution of up to $2$ photons, the success probability goes below $50\%$, which is worse than the passive linear case. The states $\ket{\psi^\pm}$ produce outputs that have at most one photon in a mode in the case of passive linear optics, whereas in our schemes, $\ket{\psi^\pm}$ produce a significant number of outputs that have more than two photons in a mode due to squeezing. Ignoring these higher-photon-number terms leads to a reduced success probability.}, which may be possible with current technology \cite{rosenberg2005, hadfield2009}.
\begin{table}[h]
\caption{Experimental Success Probabilities} % title name of the table
\centering % centering table
\setlength{\tabcolsep}{10pt}
\begin{tabular}{c c c } % creating 3 columns
\hline\hline % inserting double-line
         & PNRD & Success\\
Encoding & Upper Limit & Probability
\\[0.5ex]
\hline % inserts single-line
% Entering 1st row
& 2 Photons & 26.5\% \\
Dual-Rail & 5 Photons & 56.1\% \\
& 10 Photons & 63.2\% \\[0.5ex]
\hline
% Entering 2nd row
& 2 Photons & 41.8\% \\
Single-Rail & 5 Photons& 56.3\%\\
& 10 Photons & 62.0\% \\[0.5ex]
\hline % inserts single-line
\end{tabular}
\label{tab:ExpSuccessProb}
\end{table}

A further important experimental consideration is whether, and to what extent, our schemes are robust against photon losses (for example, through an inefficient detector). The SR scheme, as in the passive case, is not robust in this regard. But, note that in DR a passive linear setup could detect single-photon losses. Remarkably, our squeezing-enhanced DR scheme achieves the same. To see this, recall Eqs. (\ref{eq:DRPhiAfterPBS}) and the fact that squeezing adds photons in pairs. Hence, after the four single-mode squeezers, the Bell states have the form
\begin{align*}
&\ket{\psi^+}\rightarrow\sim\ket{odd,odd,even,even}+\ket{even,even,odd,odd},\\
&\ket{\psi^-}\rightarrow\sim\ket{odd,even,odd,even}-\ket{even,odd,even,odd},\\
&\ket{\phi^\pm}\rightarrow\sim\ket{even,even,even,even},
\end{align*}
where we have only shown the even/odd number of the photons in each output mode. The above equation shows that {\it measuring an even number of photons in an odd number of output modes signals a single-photon loss}. Such a measurement can be discarded, leading to a reduced success probability depending on the single-photon loss rate in the experimental setup. In contrast to the passive linear-optical case, two-photon losses cannot be reliably identified.

%%\section{Applications}
\emph{Applications---} Efficient BMs have direct applications
in quantum computation and communication. Whereas measurement-based quantum computation requires at least near-deterministic teleportation
steps \cite{gottesman1999, knill2001},
probabilistic BMs may be used in conjunction with local quantum memories for connecting the elementary segments of a quantum repeater \cite{sangouard2011}. Such a probabilistic setup is sufficient to suppress the exponential decay of entanglement due to channel transmission losses. The pair-creation rates in a general repeater chain of total length $L$
are proportional to $(L/L_0)^{\log_2(2 P_{\rm swap}/3)}$, with repeater stations
separated by distances $L_0$, and the success probability for entanglement swapping given by $P_{\rm swap}$. $P_{\rm swap}$ effectively corresponds to the BM efficiency.
This rate formula is a good approximation for small initial distribution efficiencies over $L_0$
and perfect quantum memories \cite{sangouard2011}.
%Without the use of quantum memories, in a so-called quantum relay which may still be useful
%in the presence of imperfect detectors, the rates are proportional to
%$P_{\rm swap}^{(L/L_0)-1}$. This formula works for any $P_{\rm swap}$.
For a typical repeater example
of $L=5120$ km with $L_0=20$ km, we have $L/L_0=256$. This leads to rates of $\sim 0.000152$ using optimal, passive BMs, while our local-squeezing-enhanced DR scheme
(with four squeezers at each repeater station) will ideally give $\sim 0.001140$.
Thus, in this example, we obtain an improvement of almost one order of magnitude, reducing the number of repeater chains operating in parallel from $10,000$ to $1,000$ for the distribution of at least one long-distance entangled pair per unit time. 
%which means that
%in order to distribute at least one long-distance pair (per time unit), about one thousand squeezing-enhanced repeater chains operating in parallel would suffice; in contrast to the passive scheme requiring roughly 10000 chains.

%\section{Outlook}

\emph{Summary and Outlook---} We have presented experimentally feasible schemes relying on static, active linear optics that increase the success probability of optical qubit BMs beyond 50\% without introducing any form of ancilla states. Our results open up interesting possibilities for the use of squeezing in discrete-variable quantum information processing. It remains to be seen whether the addition of vacuum modes to our schemes increases the success probability further, possibly to near 100\%.

\newpage
%\section{Acknowledgements}
\emph{Acknowledgements---} We thank T. Ralph, A. Furusawa, and T. Rudolph for helpful input and discussions. H.Z. also enjoyed talking to C. Cafaro, R. Wickert, and V. Vaibhav about various related questions, and acknowledges support from the Emmy Noether program of the DFG.

%%%%%%%%%%%%%%%%%%%%%%%%%%%%%%%%%%
% Bibliography
%%%%%%%%%%%%%%%%%%%%%%%%%%%%%%%%%%

\bibliography{UBSDSqueezingPRL}
\clearpage
\begin{widetext}
\section{Supplementary Material}
Throughout the supplement, we use commas to separate the photon numbers in different modes, e.g., what in the main text was written as $\ket{0002}$ is written as $\ket{0,0,0,2}$.

\subsection{Passive Linear Optics Upper Bound for Unambiguous Bell Measurement in Single-Rail}
The mathematical formulation presented in Ref. \cite{calsamiglia2001} (which dealt with dual-rail Bell measurements with the help of passive linear optics and vacuum ancillae) could also be applied to the case of single-rail Bell measurements. The following argument, however, is not only simpler, but also includes the possibility of conditional dynamics, which was excluded in Ref \cite{calsamiglia2001}. We show that in single-rail, passive linear networks aided by an arbitrary number of vacuum modes and conditional dynamics give a tight upper bound of 50\% on unambiguous Bell measurement success probability.  

Consider the input $\ket{\phi^\pm}\otimes\ket{\xi}$ to a passive linear circuit, where $\ket{\xi}$ is the $(n-2)$-mode vacuum ancilla, $\ket{0}^{\otimes (n-2)}$. For concreteness, the states $\ket{\phi^\pm}$ are input in modes one and two (this numbering is arbitrary). The effect of an $n$-mode passive linear circuit is to mix the input creation operators $(a_i^{\dagger})$, such that the
output operators $(b_i^{\dagger})$ are given by
\begin{equation}
b_i^{\dagger}=\sum_{j=1}^{n} U_{ij}a_j^{\dagger},
\end{equation}
where $U_{ij}$ is the $(i,j)$-element of an $n\times n$ unitary matrix. This implies that the inputs $\ket{\phi^\pm}\otimes\ket{\xi}$ are mapped to
\begin{align}
\label{eq:SRPhiPassive}
\ket{\phi^{\pm}}\otimes\ket{\xi}\rightarrow\frac{1}{\sqrt{2}}(1\pm \sum_{i,j=1}^{n} U_{1i}a_i^{\dagger}U_{2j}a_j^{\dagger})\ket{0}^{\otimes n}.
\end{align}

In the above form, it is readily seen that for a static network, the photon-number patterns in the output are identical for $\ket{\phi^+}$ and $\ket{\phi^-}$. In fact, the patterns are identical even after allowing for conditional dynamics. To see this, consider the cases of detecting zero, one or two photons in some mode $s$. If we detect two photons in $s$, then the rest of the modes contain the vacuum, which is indistinguishable for the two inputs. Now, say we detect one photon in $s$. Then the conditional state is
\begin{equation}
\ket{\phi^\pm}\otimes\ket{\xi}\rightarrow \sim \pm\sum_{k}\alpha_k a_k^{\dagger}\ket{0}^{\otimes (n-1)},
\end{equation}
where the summation over $k$ runs from mode one to $(s-1)$ and from $(s+1)$ to $n$, and $\alpha_k$ are some undetermined coefficients. Again, the output patterns are identical for the inputs even after further linear processing. To see that the case of zero photons in $s$ is also a failure event, notice that after detecting zero photons, the conditional state looks similar to that in Eq. (\ref{eq:SRPhiPassive}), except that the conditional state is over $n-1$ modes instead of $n$. This implies, once again, that any further linear processing would be futile. Hence, the probability of discriminating between the states $\ket{\phi^\pm}$ is zero. Combined with the fact that a balanced beam splitter
deterministically discriminates between the two states $\ket{\psi^\pm}$, we arrive at the tight upper bound of $50\%$ success probability for SR BM using a passive linear-optical circuit, an arbitrary number of vacuum modes and conditional dynamics.
%This is just to avoid confusion when the general expressions for the output in single-rail and dual-rail are presented below.

\subsection{Complete Output in the Dual-Rail Case}
Using Eqs. (\ref{eq:Squeezing}), we can write the effect of the squeezing operator on the states $\ket{0}$ and $\ket{2}$ as
% the effect of the squeezing operator on $\ket{0}$ and $\ket{2}$ as $\mathcal{S}\ket{2}=&\frac{\sqrt{\sech r}\tanh r}{\sqrt{2}}\sum\limits_{n=0}^\infty \frac{\sqrt{2n!}}{n!}\left(\frac{-\tanh r}{2}\right)^n\ket{2n} +\sech^{5/2} r\sum\limits_{n=1}^\infty \left(\frac{-\tanh r}{2}\right)^{n-1}\frac{\sqrt{2n!}}{\sqrt{2}(n-1)!}\ket{2n}$ and $\mathcal{S}\ket{0}=&\sqrt{\sech r}\sum\limits_{n=0}^\infty \frac{\sqrt{2n!}}{n!}\left(\frac{-\tanh r}{2}\right)^n\ket{2n}$.
\begin{align*}
\mathcal{S}\ket{0}=&\sqrt{\sech r}\sum_{n=0}^\infty \frac{\sqrt{2n!}}{n!}\left(\frac{-\tanh r}{2}\right)^n\ket{2n},\\
\mathcal{S}\ket{2}=&\frac{\sqrt{\sech r}\tanh r}{\sqrt{2}}\sum_{n=0}^\infty \frac{\sqrt{2n!}}{n!}\left(\frac{-\tanh r}{2}\right)^n\ket{2n} +\sech^{5/2} r\sum_{n=1}^\infty \left(\frac{-\tanh r}{2}\right)^{n-1}\frac{\sqrt{2n!}}{\sqrt{2}(n-1)!}\ket{2n}.
\end{align*}
%\enlargethispage{2\baselineskip}

The next step is to apply the squeezing operator on the states $\ket{\phi^\pm}$ as given in Eqs. (\ref{eq:DRPhiAfterPBS}):

\begin{align*}
\hspace*{-.5cm}\ket{\phi^\pm}\rightarrow &\frac{\tanh r}{\sqrt{2}}\sech^2 r \sum\limits_{\substack{m=0,n=0\\p=0,q=0}}^\infty \frac{\sqrt{(2m)!(2n)!(2p)!(2q)!}}{m!n!p!q!}\left(\frac{-\tanh r}{2}\right)^{m+n+p+q}(1\pm 1)\ket{2m,2n,2p,2q}\nonumber\\
&+\frac{\sech^4 r}{2}\sum_{p=0,q=0}^\infty\frac{\sqrt{(2p)!(2q)!}}{p!q!}\left(\frac{-\tanh r}{2}\right)^{p+q}\nonumber\\
&\times\left(\sum_{m=1,n=0}^\infty\left(\frac{-\tanh r}{2}\right)^{m+n-1}\frac{\sqrt{(2m)!(2n)!}}{\sqrt{2}(m-1)!n!}\ket{2m,2n,2p,2q} \pm \sum_{m=0,n=1}^\infty\frac{\sqrt{(2m)!(2n)!}}{\sqrt{2}m!(n-1)!}\left(\frac{-\tanh r}{2}\right)^{m+n-1}\ket{2m,2n,2p,2q} \right)\nonumber\\
&+\frac{\sech^4 r}{2}\sum_{m=0,n=0}^\infty\frac{\sqrt{(2m)!(2n)!}}{m!n!}\left(\frac{-\tanh r}{2}\right)^{m+n}\nonumber\\
&\times\left(\sum_{p=0,q=1}^\infty\left(\frac{-\tanh r}{2}\right)^{p+q-1}\frac{\sqrt{(2p)!(2q)!}}{\sqrt{2}p!(q-1)!}\ket{2m,2n,2p,2q} \pm \sum_{p=1,q=0}^\infty\left(\frac{-\tanh r}{2}\right)^{p+q-1}\frac{\sqrt{(2p)!(2q)!}}{\sqrt{2}q!(p-1)!}\ket{2m,2n,2p,2q} \right).
\end{align*}

Next, we want to identify the output based on the number of photons in each mode. Since this requires forming linear combinations of the above summations, the coefficients of the terms turn out to be similar with minor differences in relative signs. Hence, in order to write the complete output compactly, we define the functions $g^\pm$, $h^\pm$, $j^\pm$ and $k^\pm$ with generic mathematical operations ($\&_1$, $\&_2$, $\&_3$ and $\&_4$) as follows:

\begin{subequations}
\label{eq:CompactCoefficients}
\begin{align}
&g^\pm(m,\&_1)=\frac{\sech^2 r}{2}  \frac{\sqrt{(2m)!}}{m!} \left(\frac{-\tanh r}{2}\right)^m \left(\alpha^\pm\mbox{ }\&_1 \mbox{ }m\beta \right),\\
&h^\pm(m,n,\&_1,\&_2)=\frac{\sech^2 r}{2}\frac{\sqrt{(2m)!(2n)!}}{m!n!}\left(\frac{-\tanh r}{2}\right)^{m+n}\left(\alpha^\pm\mbox{ } \&_1 \mbox{ }m\beta \mbox{ } \&_2 \mbox{ }n\beta \right),\\
&j^\pm(m,n,p,\&_1,\&_2,\&_3)=\frac{\sech^2 r}{2}\frac{\sqrt{(2m)!(2n)!(2p)!}}{m!n!p!}\left(\frac{-\tanh r}{2}\right)^{m+n+p}\left(\alpha^\pm\mbox{ }\&_1\mbox{ }m\beta \mbox{ }\&_2\mbox{ }n\beta \mbox{ }\&_3\mbox{ }p\beta \right),\\
&k^\pm(m,n,p,q,\&_1,\&_2,\&_3,\&_4)=\frac{\sech^2 r}{2}\frac{\sqrt{(2m)!(2n)!(2p)!(2q)!}}{m!n!p!q!}\left(\frac{-\tanh r}{2}\right)^{m+n+p+q}\left(\alpha^\pm\mbox{ }\&_1\mbox{ }m\beta \mbox{ }\&_2\mbox{ }n\beta \mbox{ }\&_3\mbox{ }p\beta\mbox{ }\&_4\mbox{ }q\beta \right),
\end{align}
\end{subequations}

where $\alpha^\pm=\sqrt{2}\tanh r (1\pm 1)$ and $\beta=-\sqrt{2}\sech^2 r/\tanh r$ (note that we have defined $\alpha^\pm$ slightly differently here compared to Eq. (\ref{DRPhiFirstOrder}) in order to highlight the exponentially decaying factor of $\sech^2 r$ with each coefficient). Using these functions we can write the output from the linear network as:
\begin{align}
\label{eq:DR_Phi_Output}
&\ket{\phi^\pm}\rightarrow \frac{\sech^2 r}{2}\alpha^\pm\ket{0,0,0,0}+\sum_{m}^\infty\left[g^\pm(m,+)(\ket{2m,0,0,0}+\ket{0,0,0,2m})+g^\pm(m,\pm)(\ket{0,2m,0,0}+\ket{0,0,2m,0})\right]\nonumber\\
&+\sum_{m,n}^\infty\left[h^\pm(m,n,+,\pm)(\ket{2m,2n,0,0}+\ket{2m,0,2n,0})+h^\pm(m,n,+,+)\ket{2m,0,0,2n}+h^\pm(m,n,\pm,\pm)\ket{0,2m,2n,0}\right.\nonumber\\
&+\left. h^\pm(m,n,\pm,+)(\ket{0,2m,0,2n}+\ket{0,0,2m,2n})\right] +\sum_{m,n,p}^\infty\left[j^\pm(m,n,p,\pm,\pm,+)\ket{0,2m,2n,2p}+j^\pm(m,n,p,+,\pm,\pm)\ket{2m,2n,2p,0}\right.\nonumber\\
&+\left.j^\pm(m,n,p,+,\pm,+)(\ket{2m,0,2n,2p}+\ket{2m,2n,0,2p})\right]+\sum_{m,n,p,q}^\infty k^\pm(m,n,p,q,+,\pm,\pm,+)\ket{2m,2n,2p,2q},
\end{align}

where all the lower limits on the summations start from one. Two-photon terms are made unambiguous by setting their coefficients equal to zero for $\ket{\phi^+}$ input, i.e., by setting $g^+(1,+)=0$ (or equivalently $\alpha^++\beta=0$). Higher order distinguishable terms can be enumerated by noticing that some of the coefficients $h^-(m,n,\&_1,\&_2)$, $j^-(m,n,p,\&_1,\&_2,\&_3)$ and $k^-(m,n,p,q,\&_1,\&_2,\&_3,\&_4)$ in Eqs. (\ref{eq:CompactCoefficients}) and (\ref{eq:DR_Phi_Output}) can be made zero for suitable values of $m$, $n$, $p$ and $q$. This is succinctly presented in Table [\ref{tab:DRDistCrit}]:
\begin{table}[h]
\caption{Higher Order Unambiguous Terms in $\ket{\phi^+}$} % title name of the table
\centering % centering table
\setlength{\tabcolsep}{10pt}
\begin{tabular}{ l  l } % creating 2 columns
\hline\hline % inserting double-line
Condition & Unambiguous Terms in $\ket{\phi^+}$
\\[0.5ex]
\hline % inserts single-line
% Entering 1st row
$ m=n$ & $\ket{2m,2n,0,0},\mbox{ }\ket{2m,0,2n,0},\mbox{ } \ket{0,2m,0,2n},\mbox{ } \ket{0,0,2m,2n} $ \\
$ m-n+p=0$ & $\ket{2m,0,2n,2p}, \mbox{ }\ket{2m,2n,0,2p}$ \\
$m-n-p=0$ & $\ket{2m,2n,2p,0}$ \\
$m-n-p+q=0$ & $\ket{2m,2n,2p,2q}$ 
\\[0.5ex]
\hline % inserts single-line
\end{tabular}
\label{tab:DRDistCrit}
\end{table}

%\begin{align}
%&m-n=0\Rightarrow\ket{2m,2n,0,0} \wedge \ket{2m,0,2n,0} \in \ket{\phi^+}_{unique} &\hspace{12pt} m-n=1\Rightarrow\ket{2m,2n,0,0} \wedge \ket{2m,0,2n,0} \in \ket{\phi^+}_{unique}\nonumber\\
%&n-m=0 \Rightarrow\ket{0,2m,0,2n} \wedge \ket{0,0,2m,2n} \in \ket{\phi^+}_{unique} &n-m=1\Rightarrow\ket{0,2m,0,2n} \wedge \ket{0,0,2m,2n} \in \ket{\phi^+}_{unique} \nonumber\\
%&m-n+p=0\Rightarrow\ket{2m,0,2n,2p} \wedge \ket{2m,2n,0,2p}\in\ket{\phi^+}_{unique}&m-n-p=0\Rightarrow\ket{2m,2n,2p,0}\in\ket{\phi^+}_{unique}\nonumber\\
%&m-n-p+q=0\Rightarrow\ket{2m,2n,2p,2q}\in\ket{\phi^+}_{unique}\nonumber\\
%\end{align}

Evaluating all the unambiguous terms that give up to $0.01\%$ contribution to the success probability, we arrive at a probability of $37.49\%$ for $\ket{\phi^+}$ and $19.75\%$ for $\ket{\phi^-}$, giving us an overall success probability of $64.3\%$.

%\pagebreak
\subsection{Complete Output in the Single-Rail, Equal Squeezing Case}
For the case of equal squeezing, the complete $\ket{\phi^\pm}$ output can be written as:

\begin{align}
\label{Phi_EqualSq}
\ket{\phi^\pm}\rightarrow &\frac{\sech r}{\sqrt{2}}\left(\ket{0,0} + \sum_{n}^\infty \left(\frac{-\tanh r}{2}\right)^{n-1}\frac{\sqrt{(2n)!}}{2n(n-1)!}\left((-\tanh r\mp n\sech^2 r)\ket{2n,0} + (-\tanh r\pm n\sech^2 r)\ket{0,2n}\right) \right)\nonumber\\
&+\frac{\sech r}{\sqrt{2}}\left( \sum_{m,n}^\infty \left(\frac{-\tanh r}{2}\right)^{m+n-1} \frac{\sqrt{(2m)!(2n)!}}{2\mbox{ }m!n!}\left(-\tanh r\pm m\sech^2r\mp n\sech^2 r\right)\ket{2n,2m} \right),
\end{align}
where the lower limit on the summations again starts from one. To set the coefficient of the $\ket{02}$ term equal to zero for $\ket{\phi^+}$ input, we must satisfy the condition $-\tanh r +\sech^2 r=0$, which implies $r=0.7218$ (a squeezing of $6.2696$ dB). This condition is also fulfilled for higher photon-number states of the form $\ket{2n,2m}$ whenever $m-n=\pm 1$. For example, the states $\ket{2,4}$ and $\ket{4,6}$ unambiguously distinguish $\ket{\phi^+}$ with a probability of $1.2\%$ and $0.1\%$, respectively. Similarly, by symmetry of the above equation, the states $\ket{4,2}$ and $\ket{6,4}$ unambiguously distinguish $\ket{\phi^-}$ with a probability of $1.2\%$ and $0.1\%$, respectively. Adding up the contributions from all the terms that give at least $0.01\%$ success probability gives us an overall probability of $62.5\%$.

%For brevity, we will denote the operators $\exp{(-\tanh r_1 a^\dagger_1^2/2)}$ and $\exp{(-\tanh r_2 a^\dagger_2^2/2)}$ as $\c{F}_1$ and $\c{F}_2$, respectively.
%\begin{align}
%\mathcal{B}_2&\mathcal{S}_2\mathcal{S}_1\mathcal{B}_1\ket{\psi^\pm}=\sqrt{\frac{\sech r_1 \sech r_2}{2}}\sum_{m,n,p=0}^\infty \frac{x^my^nz^p}{m!n!p!}\nonumber\\
%&\times(\Omega^\pm\sqrt{(2n+p)!(2m+p+1)!}\ket{2m+p+1,2n+p}\nonumber\\
%&+\omega^\pm\sqrt{(2m+p)!(2n+p+1)!}\ket{2m+p,2n+p+1})
%\end{align}
%The Fock space representation of $\exp{(-\tanh r a^{\dagger^2}/2)}$ on %$\ket{0}$ and $\ket{2}$ is straightforward:
%\begin{align*}
%&\exp{(-\tanh r a^{\dagger^2}/2)}\ket{0}=\sum_{n=0}^\infty \left(\frac %{-\tanh r}{2}\right)^n  \frac{\sqrt{(2n)!}}{n!}\ket{2n}\\
%&\exp{(-\tanh r a^{\dagger^2}/2)}\ket{2}=\sum_{n=0}^\infty \left(     %\frac{-\tanh r}{2}\right)^{n}\nonumber\\
%&\times \frac{\sqrt{(2n+2)!}}{\sqrt{2} (n)!}\ket{2n+2}.
%\end{align*}
\subsection{Numerical Calculations for the Single-Rail Case}
Numerical calculations on a two-mode network with real parameters were done in Mathematica (the code is available upon request). Quantum Mathematica add-on by Jose Luis G$\acute{o}$mez-Mu$\tilde{n}$oz and Francisco Delgado was used for part of the simulation, which is  available at \url{http://homepage.cem.itesm.mx/lgomez/quantum/}. The objective was to see if the success probability of $62.5\%$ in SR is optimal within the context of a two-mode network characterized by real parameters in the Bloch-Messiah Reduction.

Eqs. (\ref{eq:PsiPMOutputSR}) and (\ref{eq:PhiPMOutputSR}) were used to numerically evaluate the success probability. We swept $\theta_1$ in steps of $0.01\pi$ radians, and the unit-less parameters $r_1$ and $r_2$ in steps of $0.05$  from $0$ to $1$ (for a maximum squeezing of $8.686$ dB). $\theta_2$ was constrained by one of the four conditions $\Omega^+=0$, $\Omega^-=0$, $\omega^+=0$ or $\omega^-=0$. Bell states were simulated to have up to $26$ photons in each mode (implying an error of less than 1\% from neglected terms in the success probability at a squeezing of $8.6859$ dB). An output that was $400$ times more likely to occur for an input state $\ket{A}$ than for any other input was considered to unambiguously identify $\ket{A}$.
%In reality, the output for $\ket{\psi^+}$ has to be compared only to that for $\ket{\psi^-}$, and the output for $\ket{\phi^+}$ to that of $\ket{\phi^-}$.
%Notice that for any of the four condition $\Omega^\pm=0$ or $\omega^\pm=0$, if $\theta_2$ is a solution then so is $\theta_2+\pi$. Under the transformation $\theta_2\rightarrow \theta_2+\pi$, $x$, $y$, $z$, $\gamma^\pm$, $\rho^\pm$ and $\zeta^\pm$ maintain their sign and magnitude, while $\Omega^\pm$ and $\omega^\pm$ maintain their relative sign and magnitude. Using this fact, we are free to disregard the solution $\theta_2+\pi$ in our numerical simulation.

Calculating the success probability for a total of $352,800$ numerical data points, it was found that, within a numerical accuracy of $1\%$, the optimal success probability was $62.5\%$, which happens for equal squeezing in the two arms of the network between $6.08$ dB and $6.51$ dB. This is shown in Fig. [\ref{fig:Fig_R1R2Equal_OmegaP}] where we plot the success probability as a function of squeezing and $\theta_1$ for equal squeezing in the two arms ($\theta_2$ stays constrained by the condition $\Omega^+=0$).  
%------------
\begin{figure}[htb]
\begin{center}
\begin{tabular}{c}
\includegraphics[width= .5 \columnwidth]{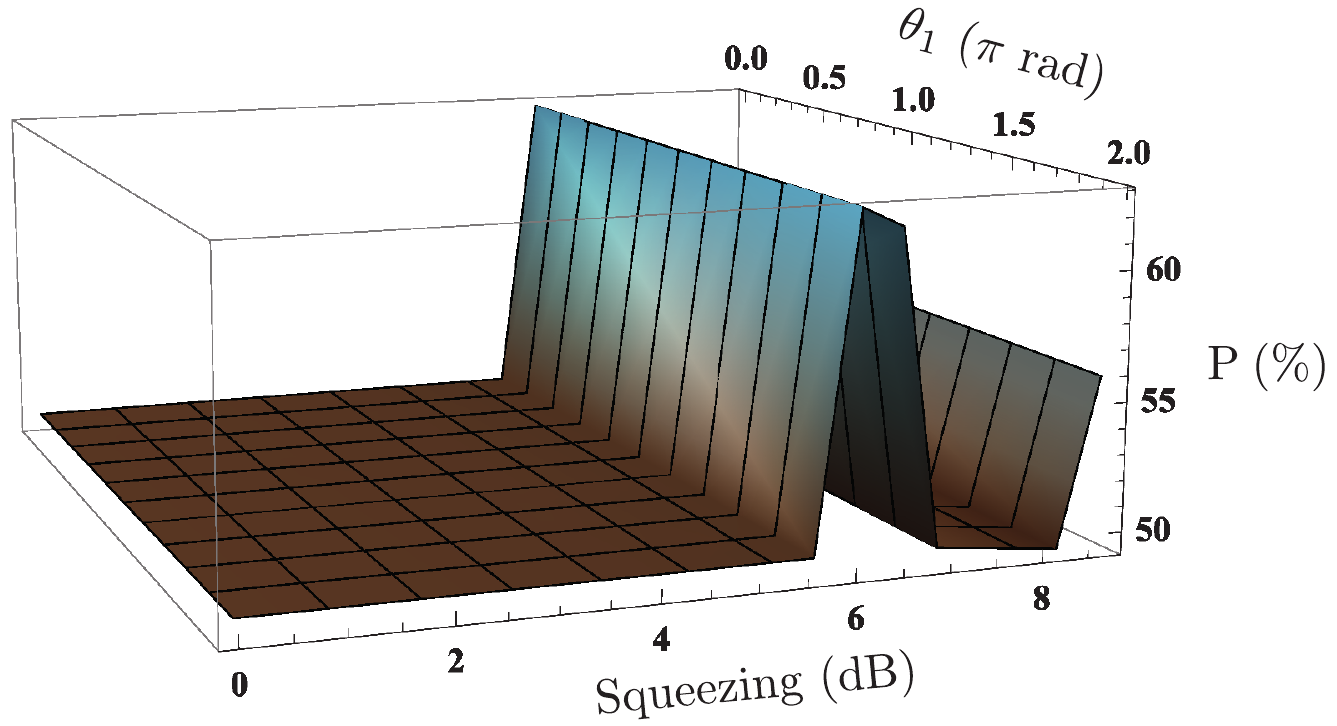}
\end{tabular}
\end{center}
\caption{Success probability (P) for $r_1=r_2$. $\theta_2$ is constrained by $\Omega^+=0$. P is bounded from below by 50\% since the states $\ket{\phi^\pm}$ are deterministically distinguishable for $r_1=r_2$. Since $\theta_2$ is constrained, every value of $\theta_1$ gives a success probability of
$62.5\%$ at a squeezing of $6.2696$ dB.}
\label{fig:Fig_R1R2Equal_OmegaP}
\end{figure}
%------------

In Fig. [\ref{fig:SuccessProbabilityPlots}], we present success probability plots as a function of squeezing in the two arms of the linear network for randomly chosen values of $\theta_1=0$, $0.67\pi$, $1.35\pi$ and $1.8\pi$ radians. These plots are all for the data set constrained by the condition $\Omega^+=0$ (the other three conditions produce similar plots). We would like to point out that the plots are not symmetric in the squeezing of the modes, since the coefficients given in Eq. (\ref{eq:PhiCoefficients}) are not symmetric in $r_1$ and $r_2$. As can be seen from the plots, qualitatively, the peak value of the success probability is about $60\%$. This can be quantitatively corroborated by looking at the success probability of individual data points. Since the plots do not add much to an intuitive understanding of the interplay between active and passive linear optics, we did not consider it worthwhile adding plots for other values of $\theta_1$.
\enlargethispage{3\baselineskip}
\begin{figure}
\centering
\mbox{\subfigure[$\mbox{ }\theta_1=0$ radians]{\includegraphics[width=2.2in]{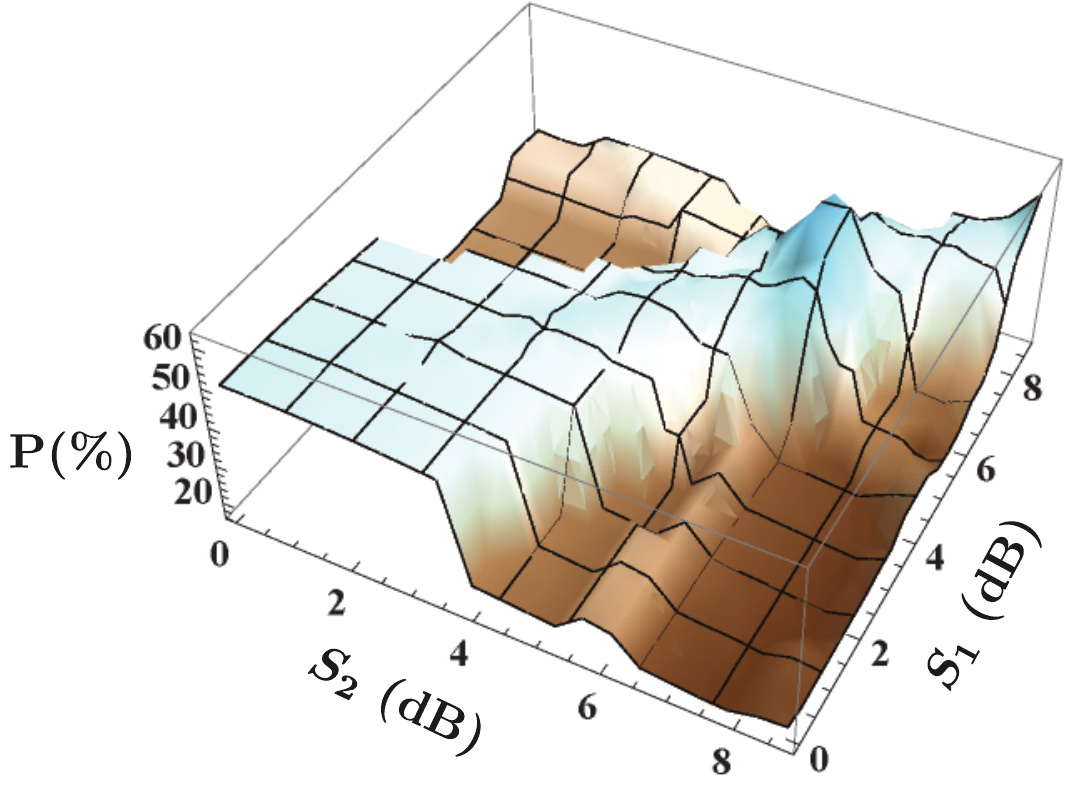}}\quad
\subfigure[$\mbox{ }\theta_1=0.67\pi$ radians]{\includegraphics[width=2.2in]{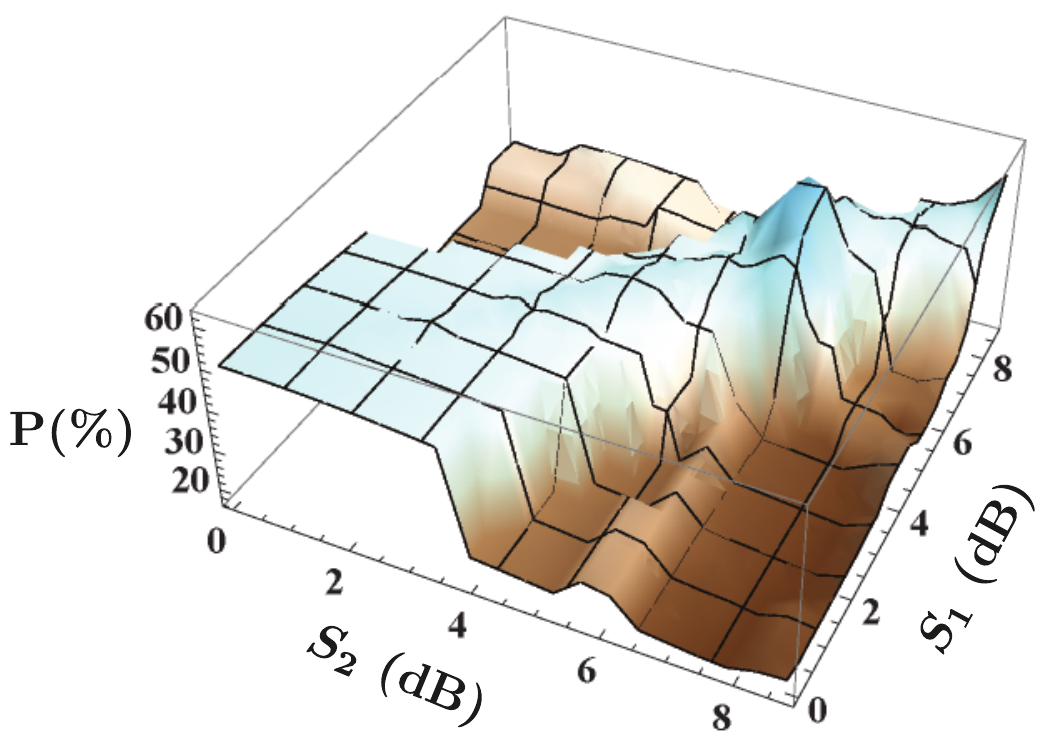} }\quad}\\
\mbox{\subfigure[$\mbox{ }\theta_1=1.35\pi$  radians]{\includegraphics[width=2.2in]{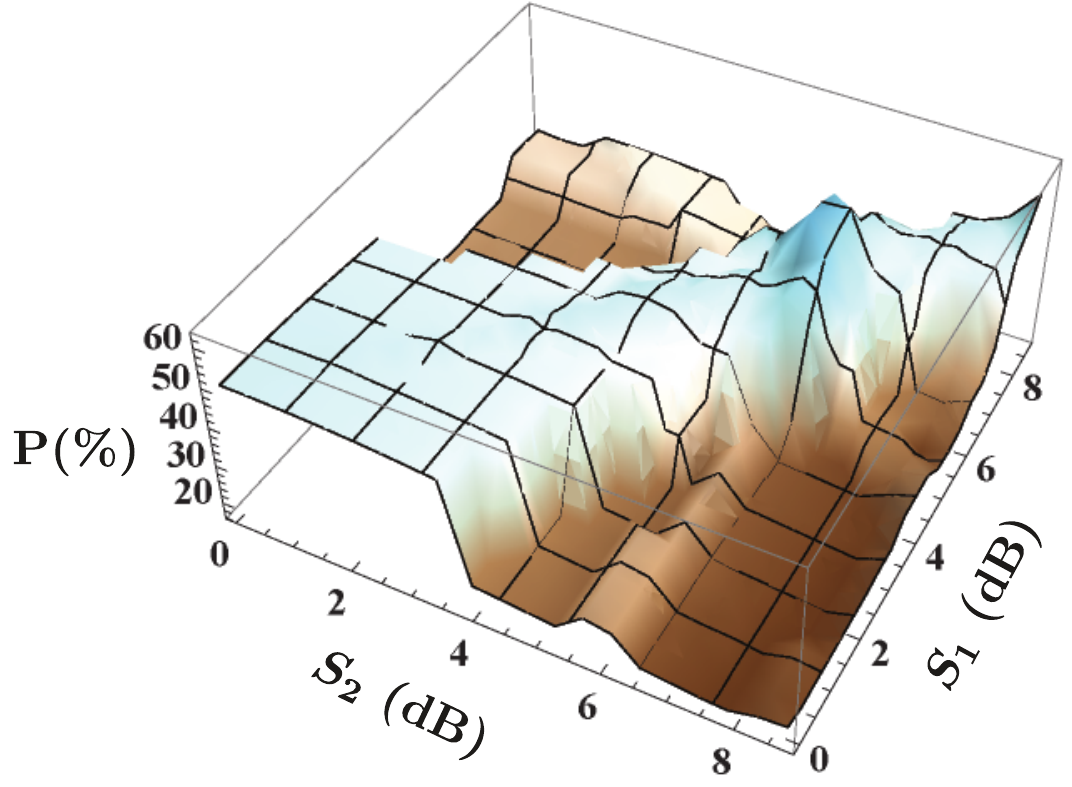} }\quad
\subfigure[$\mbox{ }\theta_1=1.8\pi$  radians]{\includegraphics[width=2.2in]{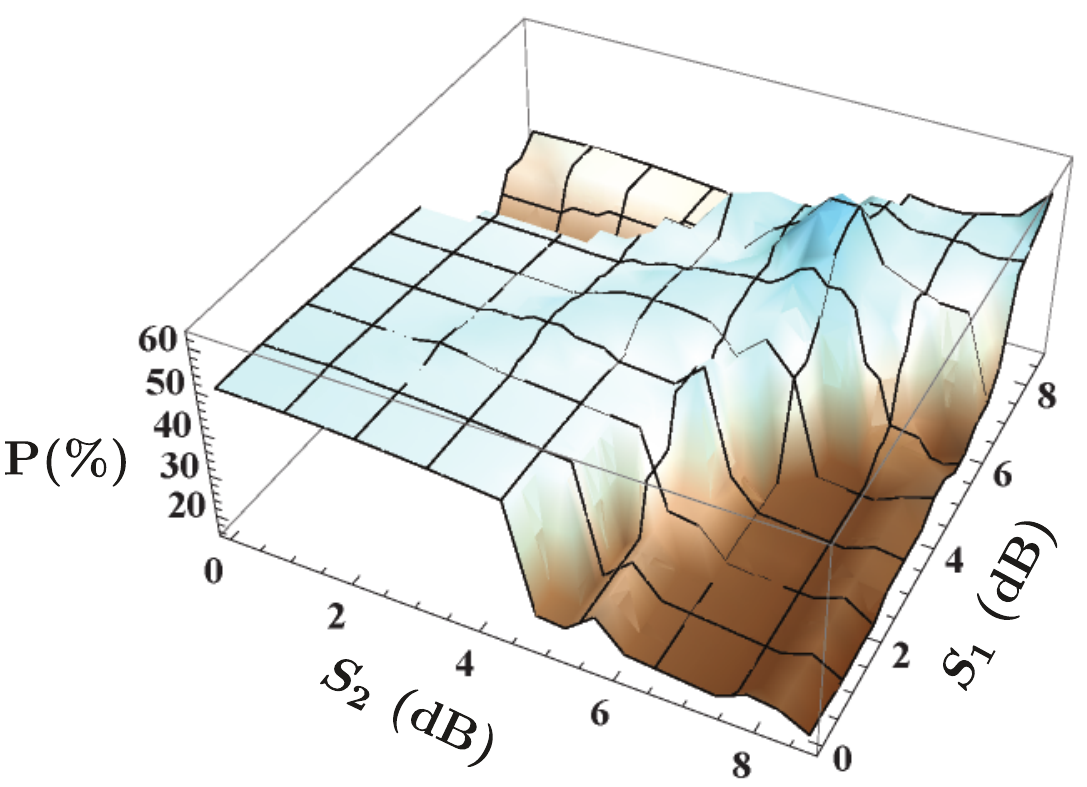} }\quad}
\caption{Success probability as a function of $r_1$ and $r_2$ for various values of $\theta_1$ after imposing the condition $\Omega^+=0$. $S_1$ and $S_2$ stand for squeezing in modes one and two of the two-mode network, respectively. For every value of $\theta_1$ we see a peak in the discrimination success probability for equal squeezing in the two modes of around $6$ dB, as expected from the analytical calculation of the equal squeezing case presented in the main text.}
\label{fig:SuccessProbabilityPlots}
\end{figure}
%------------
It is unlikely that generalizing the numerical calculations to include phase shifts in an arbitrary two-mode linear network will change the upper bound, though this will have to be addressed in future work.
\end{widetext}
%-------------
\end{document}